\documentclass{AIMS}
\usepackage{amssymb}
\usepackage{bbm}
\usepackage{graphicx}
\usepackage{hyperref}

\def\currentvolume{}
 \def\currentissue{}
  \def\currentyear{}
   \def\currentmonth{}
    \def\ppages{}

\newcommand{\pdfBoxd}[9]{\vspace{#7}
                        \begin{minipage}{#1}   
                        \centerline{\includegraphics[width= #2,angle=#9]{#3}}
                        \end{minipage}\hfill
                        \begin{minipage}{#4}
                           \centerline{\includegraphics[width= #5,angle=#9]{#6}}
                        \end{minipage}
                        \vspace{#8}}

\newcommand{\Heff}{{\sf H}^{\rm eff}}
\newcommand{\Veff}{{\sf V}^{\rm eff}}
\def\R{{\mathbbm R}}
\def\N{{\mathbbm N}}
\def\Z{{\mathbbm Z}}

\title[FPU traveling discrete breathers]{Effective Hamiltonian for traveling discrete
breathers in the FPU chain}
\author[Michael Kastner and Jacques-Alexandre Sepulchre]{}


\email{Michael.Kastner@uni-bayreuth.de}
\email{jacques-alexandre.sepulchre@inln.cnrs.fr}

\begin{document}

\maketitle

\centerline{\scshape Michael Kastner}
 \medskip

  {\footnotesize \centerline{Physikalisches Institut, Theoretische Physik I, Universit\"at
Bayreuth}
  \centerline{95440 Bayreuth, Germany} }
 \medskip

\centerline{\scshape  Jacques-Alexandre Sepulchre}
 \medskip

 {\footnotesize \centerline{Institut Non Lin\'eaire de Nice}
 \centerline{1361 Route des Lucioles, 06560 Sophia Antipolis, France} }
  \medskip


\medskip

\begin{abstract}
For the Fermi-Pasta-Ulam chain, an effective Hamiltonian is constructed, describing the
motion of approximate, weakly localized discrete breath\-ers traveling along the chain. The
velocity of these moving and localized vibrations can be estimated analytically as the group
velocity of the corresponding wave packet. The Peierls-Nabarro barrier is estimated for
strongly localized discrete breathers.
\end{abstract}

\section{Introduction}
\label{sec:intro}

Discrete breathers (DB) have the defining properties of being spatially localized and having
time-periodic dynamics. They are also called {\em intrinsically localized}, in distinction to
Anderson localization triggered by disorder. A necessary condition for their existence is the
nonlinearity of the equations of motion of the system, and the existence of discrete
breathers has been proved rigorously for some classes of systems
\cite{McKAbr,Bambusi,LiSpiMac,AKK,J,J03,JaNo}. In contrast to their analogs in continuous
systems, the existence of discrete breathers is a generic phenomenon, which accounts for
considerable interest in these objects from a physical point of view in the last decade. In
fact, recent experiments could demonstrate the existence of discrete breathers in various
real systems such as low-dimensional crystals \cite{Swanson_ea}, antiferromagnetic
materials \cite{SchwarzEnSie}, Josephson junction arrays \cite{BiUs}, molecular crystals
\cite{EdHamm}, coupled optical waveguides \cite{Mandelik_ea}, and micromechanical
cantilever arrays \cite{Sato_ea}. For a review on the topic see \cite{FlaWi}.

More than a decade ago, discrete breathers have been observed in numerical simulations of,
among others, the Fermi-Pasta-Ulam (FPU) system~\cite{ST,SPS}, a chain of nonlinearly coupled
masses. Although existence of DB as exact periodic solutions was proved already some years
ago for a large class of oscillator networks~\cite{McKAbr,SM} (not containing the FPU
system), a proof of their existence in the FPU system has been obtained only
recently~\cite{J,AKK}.

A generalization of the concept of DB are so-called traveling DB, spatially localized
solutions which travel along the chain. At least as approximate solutions they have been
observed numerically in the FPU system. A proof of their existence, however, has not been
achieved so far, and in fact it is doubted that traveling DB exist as exact solutions in this
system. The theoretical analysis of approximate traveling DB is still an open problem which
will be tackled in this article. We consider traveling DB as long-living transient structures
which can be described by an effective Hamiltonian as presented in~\cite{MS}. We distinguish
three types of localized structures, characterized by their degree of localization, which can
be found in the FPU dynamics. Examples of these different types are plotted in Figure 1,
where from type I to type III the degree of spatial localization is increasing. Type I is
similar to a standing wave. It is not a proper DB, as its localization depends on the system
size, but its energy envelope is sort of localized in a finite system. Type II is
exponentially localized, but, as the width of the structure is not that small with respect to
the intersite distance, it is called a weakly localized DB. Type III depicts a strongly
localized DB, similar to those first discovered numerically~\cite{ST}. All three of these
structures are observable in FPU systems as (transient) traveling localized objects, called
traveling discrete breathers, and we will distinguish in the following between these three
types when deriving their effective dynamics.

\begin{figure}[htb]
\centerline {
\includegraphics[width=12cm]{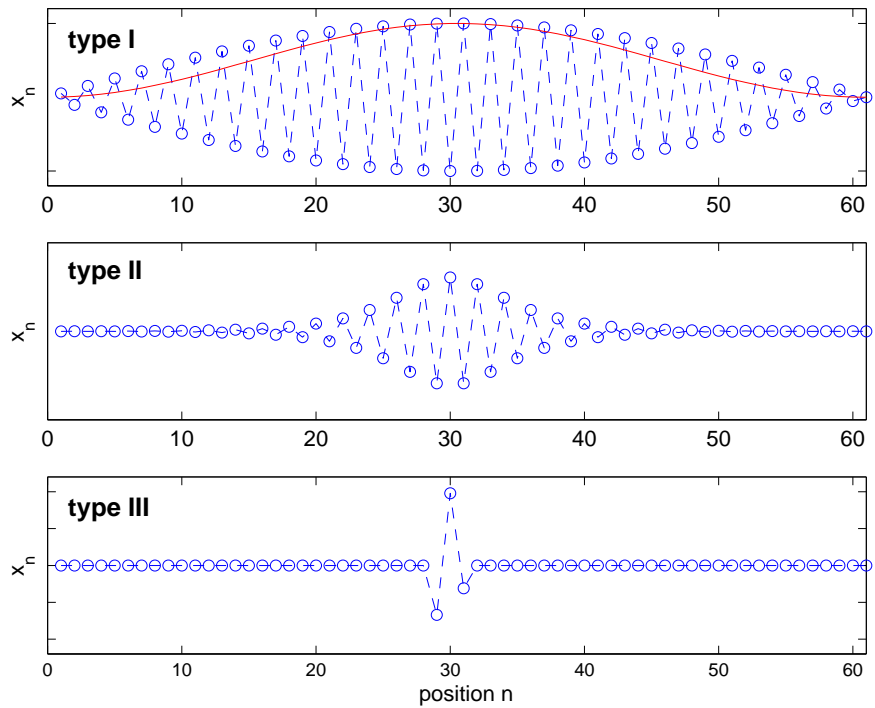}
}
\caption{ \label{f1} {\protect\small
Three types of traveling DB in FPU. Type I: weakly localized standing wave. Type II: weakly
localized DB. Type III: strongly localized DB. The red line in the upper plot represents the
local energy.}
}
\end{figure}

The outline of this article is as follows: Section~\ref{sec:FPU} provides some basic
definitions regarding the FPU model. In Section~\ref{sec:proc}, the method of the effective
Hamiltonian is briefly recalled, providing the necessary ingredients for its application.
Section~\ref{sec:waves} succinctly recalls some results from~\cite{MS}, describing the motion
of DB of type I. In Section~\ref{sec:weak}, an effective Hamiltonian is constructed for a
family of DB of type II,  providing a good framework to understand the traveling motion of
weakly localized DB in FPU chains. Finally, in Section~\ref{sec:strong}, the case of strongly
localized DB is treated. Here, instead of calculating the effective Hamiltonian explicitly,
the so-called Peierls-Nabarro barrier is computed, which provides already useful information
about the moving properties of DB of type III.

\section{FPU model}
\label{sec:FPU}

Let us consider a chain of oscillators whose dynamics is determined by the Hamiltonian
\begin{equation}
\label{eq:chainGen}
H = \sum_{n \in \Z} \left[ \frac{ p ^2 _n}{ 2 } + V(x_n) + W( x_{n+1} - x_n) \right],
\end{equation}
with momenta $p_n\in\R$ and positions $x_n\in\R$.  $V$ and $W$ characterize, respectively,
the on-site potential and the (nearest-neighbor) interaction potential. The FPU model is
defined by the choices
\begin{equation}\label{eq:W}
V = 0 \qquad\mbox{and}\qquad W (x) = \frac{k}{2} x ^2 + \frac{ \alpha }{ 3 } x ^3
+ \frac{ \beta }{ 4 } x ^4,
\end{equation}
where $k,\alpha,\beta\in\R$. Note, however, that the theory of the next section
applies to a more general class of models.

\section{Method of the effective Hamiltonian}
\label{sec:proc}

The method of the effective Hamiltonian has been exposed in \cite{MS} in a general framework.
It is briefly recalled here before applying it to the FPU model in the subsequent section.

Let us define a traveling DB of~(\ref{eq:chainGen}) as a function of the form
\begin{equation}
\label{eq:TDB}
x _n (t) = u (A, n-c t , \tilde{\omega} t),
\end{equation}
where the function $ u(A,r,s) $ is increasing in $A$, and such that $u(0,\cdot,\cdot) = 0$.
It is localized in $r$ (e.g.,  $| u(\cdot,r,\cdot)| < {\mathrm e}^{- \lambda | r |}$ for some
$ \lambda > 0$) and periodic in $s$, $u(\cdot,\cdot, s+2 \pi) = u(\cdot,\cdot,s)$. To be
slightly more general, one may think of $ct$ being replaced by a function $Q(t)$ monotonic in
$t$, where $\dot{Q} = c+f(t)$ and $f$ is a function of small amplitude, periodic with a
frequency which is small compared to $\tilde{\omega}$. The point is that, in a moving frame
with velocity $\dot{Q}$, one sees a discrete breather at rest with frequency $\tilde{\omega}$.

It is generally believed that solutions of the form~(\ref{eq:TDB}) do not exist as exact
solutions in generic systems, although a proof of this conjecture has been accomplished only
for a certain class of systems \cite{BeMcKRo}. However, it may be possible to get {\em
approximate} traveling DB of the form~(\ref{eq:TDB}) when the following conditions can be met:

\begin{enumerate}

\item Find a two-parameter family of approximate (non-traveling) discrete breath\-ers
of~(\ref{eq:chainGen}) of the form
\begin{equation}
\label{eq:famDB}
x _n (t) = u (A, n-Q, \omega t),
\end{equation}
where $A$ characterizes the amplitude of the DB and $Q$ its ``position". Typically, for a
chain having some translation symmetry, and for fixed $A$, there are two such positions where
there is an exact DB: a ``site-centered" DB with integer $Q$, and a ``bond-centered" DB with
half-integer $Q$. By interpolating between these exact solutions, approximate DB can be
constructed for any real $Q$.

Next, consider the family of functions
\begin{equation}
\label{eq:famMDB}
x _n ^{(A,Q,k)} = u (A, n-Q, \omega t-kn).
\end{equation}
 The idea is that there may be a ``Doppler effect'' of the traveling DB, $ \tilde{\omega} -
\omega = k c$, which can be described by adding a phase shift $kn$ to $\omega t$, where $k$
is called the {\em momentum}\/ of the DB. The point is that (\ref{eq:famMDB}) still describes
approximate DB for small enough $k$.

\item Compute, for each member of the family $x _n^{(A,Q,k)}$, the symplectic area
\begin{equation}
\label{eq:areaDB}
a = \int_0 ^T \left( \sum_{n} p_n^{(A,Q,k)} \dot{x}_n^{(A,Q,k)} \right) {\mathrm d}t
\end{equation}
with period $ T = 2 \pi / \omega $.
In the case where $p=\dot{x}$, like for the Hamiltonian~(\ref{eq:chainGen}), one can write
\[
a = 2 T \left\langle E_{\mbox{\rm \tiny kin}} (A,k,Q) \right\rangle,
\]
where $\langle\cdot\rangle$ denotes the mean value over one period and $E_{\mbox{\rm \tiny
kin}}$ is the kinetic energy. This calculation should result in a symplectic area $a$
proportional to $A$, or related to $A$ by a simple functional relation. If this is not the
case, the parametrization in~(\ref{eq:famMDB}) should be changed, indexing the family in
terms of $(a,Q,k) $. The fact that $a$ is a proper choice as a parameter is explained
in~\cite{McK,MS,S}. In short, this can be justified by stating that $a$ is an adiabatic
invariant of the dynamics~\cite{ArKN}.

\item Compute the symplectic form $ \sum_{n} {\mathrm d}p_n \wedge {\mathrm d}x_n$ restricted
to the coordinates $ (k,Q) $. This is achieved by evaluating
 \begin{equation}
\label{eq:taukQ}
 \tau_{kQ} = \frac{ 1 }{ T } \int_0^T \left( \sum_{n} \frac{\partial p_n }{\partial k}
\frac{\partial x_n}{\partial Q} - \frac{\partial p_n}{\partial Q} \frac{\partial
x_n}{\partial k} \right) {\mathrm d}t,
\end{equation}
which should be non-zero for any $(k,Q)$, and the restricted symplectic form reads
$\tau_{kQ}\, {\mathrm d}k \wedge {\mathrm d}Q $.

\item Then, an effective Hamiltonian is computed as the mean (over one period) of the
original Hamiltonian along the family of DB,
\begin{equation}
\label{eq:HeffDB}
  \Heff_a (k,Q) = \frac{ 1 }{ T} \int_0 ^T H \left( \left\{ x
_n^{(a,Q,k)} , p _n^{(a,Q,k)} \right\} \right) {\mathrm d}t,
\end{equation}
and, finally, the dynamics of $(k,Q)$ is given by
\begin{eqnarray}
\label{eq:effdyn}
\dot{k} & = & - \frac{ 1 }{ \tau _{kQ} } \frac{\partial \Heff_a
}{\partial Q},\\
\dot{Q} & = & \frac{ 1 }{ \tau _{kQ}} \frac{\partial \Heff_a }{\partial
k}.
\end{eqnarray}
For small enough $k$, and assuming $\partial_k \Heff_a(k,Q) = 0$ for $k=0$  (otherwise shift
$k \rightarrow k-k_0$  to have this) one may expand the effective Hamiltonian so that it
takes on the form
\begin{equation}
\label{eq:HeffDB2}
\Heff_a (k,Q) \simeq  \frac{k^2}{2 \,m(Q)} + \Veff(Q),
\end{equation}
providing that $m(Q)$ be non-zero and finite.  The latter defines an effective inertia,
although $m(Q)$ may be negative. In fact, as discussed in~\cite{AMS} (especially in examples
of Section 6 therein), the sign of $m(Q)$ can be interpreted as  the sign of an effective
anharmonicity of the breather solutions. The second term in~(\ref{eq:HeffDB2})  is a periodic
function (of period $1$) which defines a {\em Peierls-Nabarro potential} for discrete
breathers, as introduced and discussed in~\cite{AMS,MS,McK}. In these papers it is shown
that the critical points of $\Veff(Q)$ correspond to exact discrete breathers of the full
model. In the simplest situation there are two types of critical points associated,
respectively, with the stable discrete breathers and the unstable ones. The difference of
$\Veff$ at these points, say $Q^{(u)}$ and $Q^{(s)}$, is called the {\em Peierls-Nabarro
barrier}
\begin{equation}
\label{eq:E_PN}
\Delta E_{\text{\tiny PN}} = | \Veff(Q^{(u)}) - \Veff(Q^{(s)}) | .
\end{equation}
The absolute value in this equation is necessary in case of negative $m(Q)$. For $m(Q)$
positive, the minima are associated with stable DB, and the maxima with unstable DB,
otherwise the stability is just reversed. In any case, the smaller this barrier is, the
higher is the mobility of a traveling discrete breather, and the better it is in general
described by the effective Hamiltonian.

\end{enumerate}

In the next sections we apply this scheme to the analysis of the mobility of various types of
localized solutions in the FPU model.

\section{Traveling localized waves (type I)}
\label{sec:waves}

In this section, a particular type of localized solutions is considered, termed type I in the
introduction, which have the shape of a standing wave. They do not constitute proper DB, as
the localization depends on the system size. The method of the effective Hamiltonian was
applied to this type of solutions in~\cite{MS}. Here, we summarize these results in order to
compare them to those for the proper DB solutions obtained in the following sections.

Consider an FPU chain with, say, an odd number $ M = 2 N+1 $, $N\in\N$, of masses and
periodic boundary conditions. There exists, in the linear regime ($ \alpha = \beta = 0 $), a
family of standing waves
\begin{equation}
\label{eq:standingW}
x _n (t) = \frac{ 2 }{ \sqrt{ M \omega }} \sqrt{ A }\, (-1)^n \sin \left( \frac{ \pi }{ M } n
- Q \right) \cos (\omega t)
\end{equation}
whose envelope is localized in space, as it is a sine function with wavelength equal to twice
the chain length (see Figure~\ref{f1}, type I). Moreover, it is observed numerically that the
localization strength is enforced when the nonlinearity is turned on, $ \alpha,\beta \neq 0
$. This family (\ref{eq:standingW}) complies with (\ref{eq:famDB}), as it is parametrized by
$A$ and $Q$, while $ \omega = 2 \sin \left( \frac{ N \pi }{ M}\right) $ is fixed. Such a wave
can be put into slow motion when $\alpha$ and $\beta$ are non-zero.

To ``add momentum'' to this standing wave, it is convenient not to follow (\ref{eq:famMDB}),
but rather to consider the expression
\begin{equation}
\label{eq:sWk}
x _n (t) = \frac{ 1 }{ \sqrt{ M \omega }} \left[ \sqrt{ A +k } \sin \left( \omega t
- \sigma n + Q \right)- \sqrt{  A - k } \sin \left(
\omega t + \sigma n - Q \right)  \right],
\end{equation}
where $\sigma = \frac{ 2\pi N}{ 2N+1 }$ and $k$ is taken as the momentum parameter conjugate to $Q$. The motivation
for
favoring
this parametrization is related to the structure of the linearized FPU chain and has been
explained in~\cite{MS}.

\begin{figure}[hbt]
\vspace{-0cm}
\pdfBoxd{6cm}{6cm}{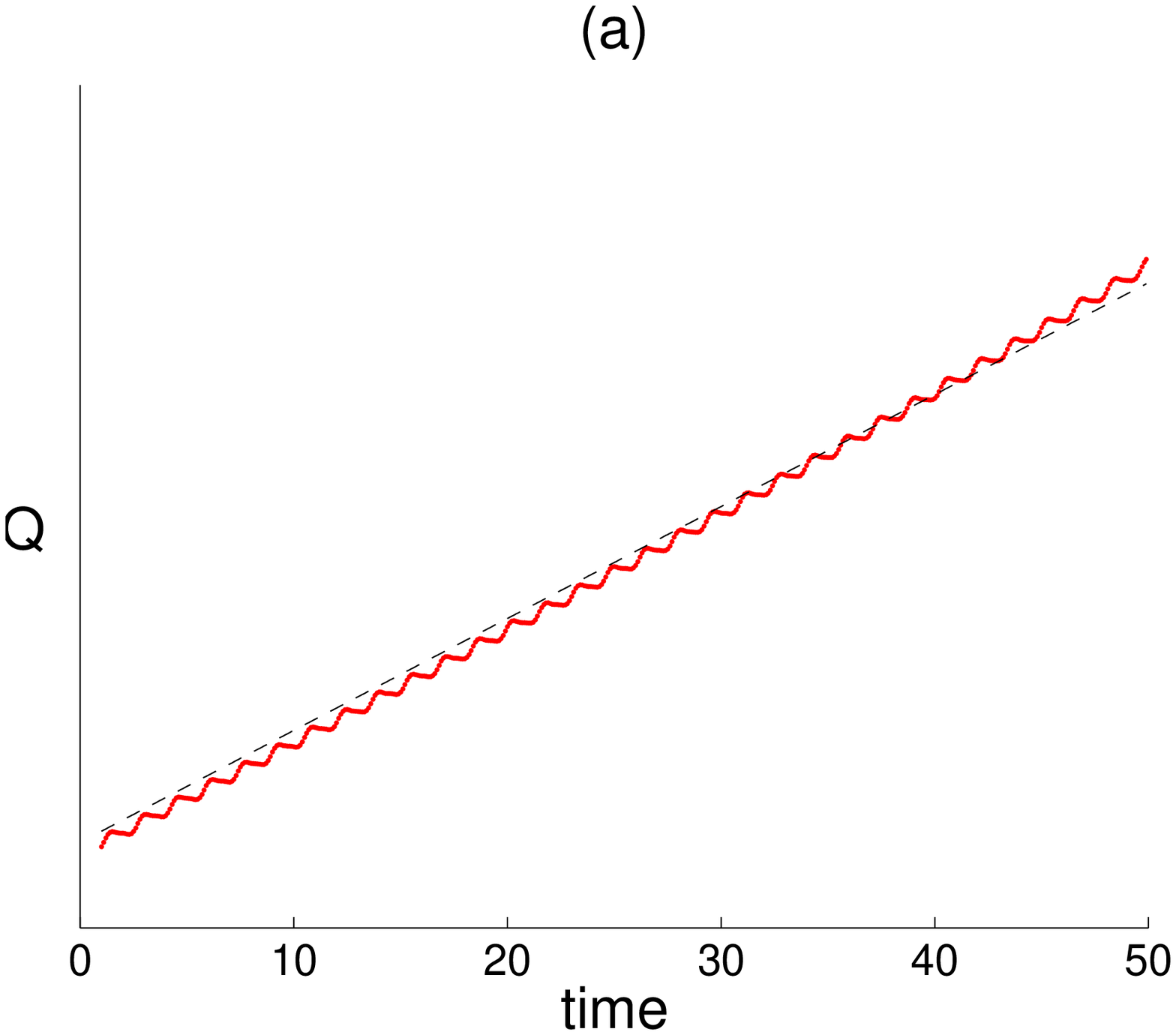}{6cm}{6cm}{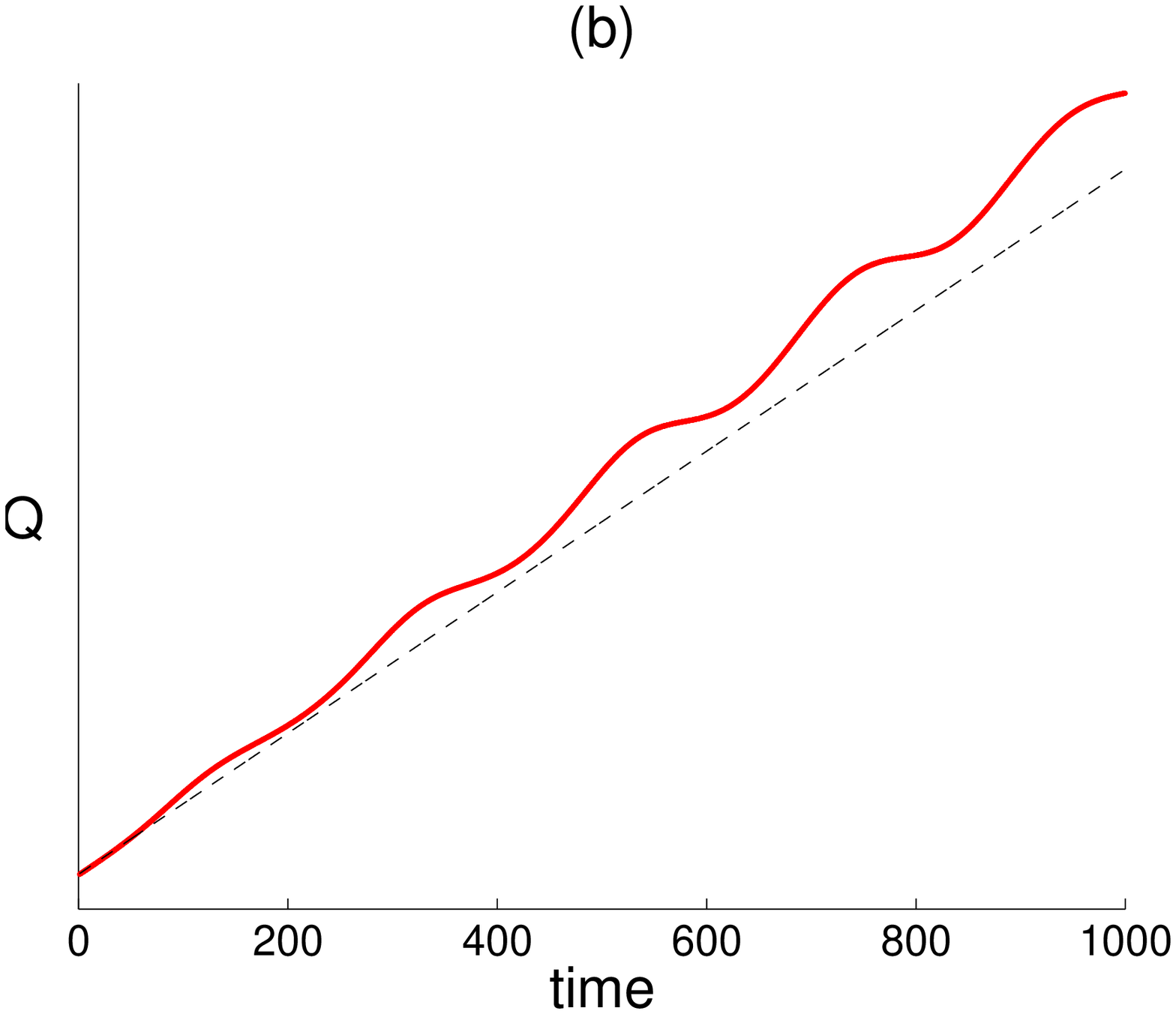}{0ex}{-0cm}{0}
\caption{\label{Fig:thetat}
Time evolution of the coordinate $Q$. Straight lines indicate the predictions of the
effective dynamics (\ref{eq:drift}). The parameters of the FPU model in (\ref{eq:W}) are
chosen as $\alpha = 0$, $\beta = 1$, $A = 0.05$, $Q_0 = -\frac{\pi}{2}$, and $k_0 = -0.01$.
Panel (a) is an enlargement of a part of panel (b), corresponding to the time lapse $[0,50]$.
}
\end{figure}

With these settings, the symplectic area~(\ref{eq:areaDB}) can be computed, yielding $a = A/
2 \pi $. The computation of $\tau _{kQ} $ from Eq.~(\ref{eq:taukQ}) gives $\tau _{kQ} =1$,
and so the effective symplectic form corresponding to the coordinates $(k,Q)$ is simply
canonical. Finally, one computes the averaged Hamiltonian (\ref{eq:HeffDB}), obtaining
\begin{equation}
\label{eq:sWHeff}
\Heff_A(k) \simeq \omega  A + \frac{3}{16} \beta \frac{\omega^2}{M} \left(3 A^2 - k^2\right),
\end{equation}
from which the equations of motion
\begin{equation}
\label{eq:drift}
\dot{Q} = -\frac{3}{8} \beta \frac{\omega^2}{M} k t\qquad\mbox{and}\qquad\dot{k} = 0
\end{equation}
are deduced. Therefore, to this order of approximation, there is no Peierls-Nabarro barrier,
and the motion of a traveling localized wave (\ref{eq:standingW}) is a slow drift at constant
speed $M \dot{Q} / \pi$.\footnote{Note that the speed depends on the system size.} This
result is confirmed by direct numerical simulation of the FPU system in the weakly nonlinear
regime, using~(\ref{eq:sWk}) as initial condition and plotting the variable $Q$ versus time
$t$ (Figure~\ref{Fig:thetat}). The prediction of (\ref{eq:drift}) turns out to be valid for
small times $t$. The discrepancy encountered for larger $t$ might be reduced by considering
higher order terms in (\ref{eq:sWHeff}).

\section{Weakly localized discrete breathers (type II)}
\label{sec:weak}

The first proof of the existence of DB in FPU type systems is due to Livi, Spicci, and MacKay \cite{LiSpiMac} for a
diatomic chain of alternating masses $m$ and $M$ with large mass ratio $m/M$. Existence of DB in monoatomic FPU
systems [as defined in (\ref{eq:chainGen}) and (\ref{eq:W})] has been proved only recently, either using a
variational
approach~\cite{AKK}, or elaborating a technique of central manifold reduction~\cite{J,J03}.
The latter existence result obtained by James will be used as a starting point in order to
construct a family of DB of the form (\ref{eq:famDB}). In his formulation~\cite{J,J03}, James
works with the force variable
\begin{equation}
\label{eq:y}
y _n = W ^{ \prime } (x _n - x _{n-1} )
\end{equation}
which is one-to-one related to the displacements $x_n$, provided that $W^{\prime}$ is
invertible.
Suppose
\begin{equation}
W^{ \prime \prime } (0) = \frac{\omega _0 ^2}{4} = 1.
\end{equation}
Then the main result of~\cite{J} can be formulated as follows: In the FPU system, for $\omega
- \omega _0 > 0$ sufficiently small, there exist DB of the form
\begin{equation}
\label{eq:yn}
y _n \simeq (-1) ^n u _n \, \cos (\omega t) + \mathcal{O}(\omega-\omega_0),
\end{equation}
 where $\mathcal{O}$ denotes the Landau order symbol, and $u_n$ satisfies the recurrence
relation
\begin{equation}
\label{eq:recur}
u _{n+1} + u _{n-1} - 2 u _n = \kappa ^2 u_n - \gamma u _n ^3 + \mbox{\rm h.o.t.}
\end{equation}
with the condition $ | u_n | \rightarrow 0 $ as $ n \rightarrow \pm \infty$. Higher oder
terms (h.o.t.) have been neglected in (\ref{eq:recur}). Equivalently, $ u _n $ (with $ n \in
\Z$) must be a homoclinic orbit of the two-dimensional map $(u _n , u _{n-1} )
\rightarrow (u _{n+1} , u _n )$ defined in (\ref{eq:recur}). In this map, the parameters are
defined as
\begin{equation} \label{eq:kappagamma}
 \kappa^2 = \omega^2 - 4 \qquad\mbox{and}\qquad \gamma = \frac{3\beta}{\omega_0^2} - \left(
\frac{2 \alpha}{\omega_0^2} \right) ^2,
\end{equation}
where $\gamma$ has to be positive in order to guarantee existence of DB. So the assumption $
\omega ^2 \gtrapprox \omega _0 ^2 =4$ implies that $ \kappa ^2 \ll 1 $
in~(\ref{eq:kappagamma}). In addition, let us suppose that $\kappa ^2 \sim \gamma |u _n| ^2$.
Then, (\ref{eq:recur}) can be approximated by the differential equation
\begin{equation}
\label{eq:ode_v}
v ^{ \prime \prime } = v - v ^3
\end{equation}
with $ u_n = \frac{ \kappa }{ \sqrt{ \gamma }} v ( n \kappa )$. Equation~(\ref{eq:ode_v}) has
a family of solutions homoclinic to $ v = 0 $, namely
\begin{equation}
\label{eq:homoclinic}
v _q ( \xi ) = \frac{ \sqrt{ 2 } }{ \cosh [\kappa ( \xi - q) ] }
\end{equation}
parametrized by $q \in \R$. Hence, one deduces an approximate analytic expression
for~(\ref{eq:yn}), reading
\begin{equation}
\label{eq:analytic_yn0}
y _n (t) \simeq (-1) ^n \kappa \sqrt{ \frac{ 2 }{ \gamma }} \frac{ \cos (\omega t ) }{ \cosh
[\kappa (n - Q ) ] }
\end{equation}
which represents a family of approximate DB of the form~(\ref{eq:famDB}). As an example, a
member of this family is plotted in Figure~\ref{f1} (type II) for $\kappa = 0.2$. Finally,
substituting $\omega t$ by $\omega t - kn $ in the cosine function with $k\approx\pi$,
provides the family of functions
\begin{equation}
\label{eq:analytic_yn}
y _n (t) \simeq  \kappa \sqrt{ \frac{ 2 }{ \gamma }} \frac{ \cos (\omega t -kn) }{ \cosh
[\kappa (n - Q ) ] }
\end{equation}
of the form (\ref{eq:famMDB}) which can serve as a good starting point for the construction
of an effective Hamiltonian for (approximate) traveling DB. Equation (\ref{eq:analytic_yn0})
is recovered for $k=\pi$. The family (\ref{eq:analytic_yn}) is indexed by three independent
parameters, $Q \in \R $, $ |k-\pi | \ll 1$, and $ \kappa \ll 1 $.

\subsection{Change of coordinates}

The Hamiltonian~(\ref{eq:chainGen}) is written in canonical coordinates $(x_n,p_n)$.  It
turns out to be convenient to work in relative coordinates $(\rho_n,J_n)$ defined by
\begin{eqnarray}
\rho _n & = & x _n - x _{n-1}, \nonumber \\
p _n & = & J _n - J _{n+1}, \label{eq:rhoJ}
\end{eqnarray}
and in the following we will rewrite the FPU Hamiltonian [(\ref{eq:chainGen}) with $V = 0$]
as well as the family of DB (\ref{eq:analytic_yn}) in these coordinates. We assume the
original coordinates to satisfy the conditions
\begin{eqnarray}
 \sum_{n } | p _n | & < & \infty, \nonumber \\
 \sum_{n } | x _n - x _{n-1} | & < & \infty, \nonumber \\
\lim_{n \rightarrow - \infty} x _n & = & 0. \label{eq:cond_xp}
\end{eqnarray}
For the new variables, these relations imply
\begin{eqnarray}
 \sum_{n } | J _n -J _{n+1} | & < & \infty, \nonumber \\
 \sum_{n } | \rho _n | & < & \infty. \label{eq:}
\end{eqnarray}
 In addition, it will be assumed that $\lim_{n \rightarrow - \infty} J _n = 0$. Then the
change of variables~(\ref{eq:rhoJ}) is invertible. For example, for given $\{ ( \rho _n , J
_n ) \}_{n \in \Z}$, one can compute
\begin{equation}
\label{eq:xn}
p _n = J _n - J _{n+1} \qquad\mbox{and}\qquad x _n = \sum_{k=-\infty}^{n} \rho _k.
\end{equation}
Moreover, the transformation is canonical since
\begin{eqnarray}
 \sum_{n} dp _n \wedge dx _n & = & \sum_{n} \left({\mathrm d}J _n \wedge {\mathrm d}x _{n-1}
+ {\mathrm d}J _n \wedge {\mathrm d} \rho _n - {\mathrm d}J _{n+1} \wedge {\mathrm d}x _n
\right) \nonumber \\
 & = & \sum_{n} {\mathrm d}J _n \wedge {\mathrm d} \rho_n.
\end{eqnarray}
In the new coordinates, the FPU Hamiltonian takes on the form
\begin{equation}
\label{eq:H_Jrho}
H = \sum _n \frac{1}{2} (J _n - J _{n+1} ) ^2 + W(\rho _n ),
\end{equation}
and the canonical equations read
\begin{eqnarray}
 \dot{\rho} _n & = & 2 J _n - J _{n+1} - J _{n-1}, \label{eq:rhodot} \\
\dot{J} _n & = & - W ^{ \prime } (\rho _n), \label{eq:Jdot}
\end{eqnarray}
or, equivalently,
\begin{equation}
\label{eq:FPUrho}
\ddot{\rho} _n = W ^{ \prime } (\rho _{n+1} ) + W ^{ \prime } (\rho _{n-1} ) - 2 W ^{ \prime
} (\rho _n).
\end{equation}
In order to compute an effective Hamiltonian from~(\ref{eq:H_Jrho}) we need to express the
family of DB~(\ref{eq:analytic_yn}) in terms of the variables $(\rho _n , J _n ) $. This is
achieved by considering the simple approximation
\begin{equation}
\label{eq:rho_y}
\rho _n  = (W ^{ \prime } ) ^{ -1} (y_n ) \simeq y _n + \mathcal{O}\left( y _n ^2 \right),
\end{equation}
which is valid due to $y _n $ being small. The corresponding momentum $J_n$ can be deduced
from~(\ref{eq:Jdot}), yielding
\begin{eqnarray}
J _n (t) & = & - \int_0 ^t W ^{ \prime } (\rho _n(s)) ds + J _n (0) \nonumber \\
 & = & - \int_0 ^t y _n (s) ds + J _n (0)  \nonumber \\
& = &  \frac{ \kappa }{ \omega } \sqrt{ \frac{ 2 }{ \gamma }} \frac{ \sin (\omega t - k n )
}{ \cosh [\kappa ( n - Q ) ] } \label{eq:Jn}
\end{eqnarray}
where $J _n (0)= -\frac{ \kappa }{ \omega } \sqrt{ \frac{ 2 }{ \gamma }} \frac{ \sin ( k n )
}{ \cosh [\kappa ( n - Q ) ] } $  has been chosen.  Note that assuming $(\rho _n , J _n ) $
to be small does not imply that Ansatz~(\ref{eq:analytic_yn}) can be considered as a
wave-packet obeying the linearized equations of motion. In fact, the linear limit $ \gamma
\rightarrow 0 $ renders (\ref{eq:analytic_yn}) singular.

\subsection{Effective Hamiltonian for weakly localized DB}

In this section, we show that the effective dynamics of approximate traveling DB
(\ref{eq:analytic_yn}) can be described by the Hamiltonian
\begin{equation}
\label{eq:Heff}
\Heff (k,Q) = \frac{1}{2} \frac{ \kappa }{ \gamma } (3- \cos k ) + \mathcal{O} \left( \kappa
^2 \right).
\end{equation}
This result is obtained by following the procedure recalled in Section~\ref{sec:proc}: First,
compute the symplectic area
\begin{equation}
\label{eq:areaJrho}
a = \int_0 ^T \left(\sum_n J_n \dot{\rho} _n \right)\, {\mathrm d}t.
\end{equation}
Using (\ref{eq:rhodot}), this expression can be written in terms of $ J _n $ only, reading
\begin{equation}
\label{eq:areaJ}
a = 2 \int _0 ^T \sum _n \left( J _n ^2 - J _n J _{n+1} \right){\mathrm d}t.
\end{equation}
Substituting (\ref{eq:Jn}) in this equation yields
\begin{equation}
  \label{eq:areakQ}
  a = \frac{ 4 \pi \kappa ^2 }{ \omega ^3 \gamma } \sum _n \left\{ \frac{ 1 }{ \cosh ^2
[\kappa ( n - Q )] } -  \frac{ \cos k }{ \cosh [ \kappa ( n+1 - Q)] \cosh [ \kappa ( n-Q)] }
\right\}.
\end{equation}
This series can be estimated making use of the Poisson summation formula
\begin{equation}
  \label{eq:PoissonSum}
  \sum _n f ( n- Q ) = \sum _m F(m) e ^{2 \pi{\mathrm i} m Q}
\end{equation}
where the coefficients $F(m)$ of the Fourier series are  the integer sampling of the
 Fourier transform
\begin{equation}
  F(\nu ) = \int_{-\infty}^{+\infty} f(x) e ^{-2 \pi{\mathrm i} \nu x} {\mathrm d}x
\end{equation}
The advantage of this formula is that it is sufficient to keep only a few terms in the
Fourier series~(\ref{eq:PoissonSum}) when $F(m)$ decreases rapidly for $m\to\pm\infty$.
Here one obtains
\begin{equation}\label{eq:Pois1}
 \sum _n \frac{ 1 }{ \cosh ^2 [ \kappa ( n - Q )] } = \frac{ 2 }{ \kappa } + \frac{ 4 \pi ^2
}{  \kappa ^2 \sinh \left( \frac{ \pi ^2 }{ \kappa } \right)} \cos (2 \pi Q ) + \mathcal{O}
\left[ {\textstyle \exp\left(-\frac{ 4 \pi ^2 }{\kappa}\right) } \right]
\end{equation}
and
\begin{equation}
 \sum _n \frac{ 1}{ \cosh [ \kappa ( n+1 - Q) ] \cosh [ \kappa ( n-Q)] } = \frac{ 2 }{ \sinh
\kappa }.
\end{equation}
For small $\kappa$, which corresponds to our assumption of $\omega-\omega_0$ being small, the
second term in (\ref{eq:Pois1}) is exponentially small compared to the first one. Neglecting
this small term and using~(\ref{eq:kappagamma}), the area reads
\begin{equation}
\label{eq:area1}
a = \frac{ 8 \pi \kappa }{ \gamma (4 + \kappa ^2 ) ^{3/2} } \left( 1 - \frac{ \kappa \cos k
}{ \sinh \kappa } \right) + \mathcal{O} \left[{\textstyle \exp\left(-\frac{\pi^2 }{ \kappa
}\right) } \right].
\end{equation}
 This result indicates that Ansatz~(\ref{eq:analytic_yn}) does not define a family of DB
whose area $a$ is independent of $k$. Consequently, following Section~\ref{sec:proc}, one
should in principle reparametrize this family by considering the change of variables
$(\kappa, k,Q) \rightarrow (a,k,Q) $. On the other hand, and for the sake of simplicity, one
may assume $k\approx\pi$ and, again, $\kappa$ small, to obtain
\begin{equation}
\label{eq:area2}
a = 2 \pi \frac{ \kappa }{ \gamma } + \mathcal{O} \left(\kappa ^2 , | \pi - k |^2 \right)
\end{equation}
at lowest order. In this case, $a$ is proportional to $\kappa$ and no change of
parametrization is needed.

Next, following Section~\ref{sec:proc}, the restriction $ \tau_{kQ}\, {\mathrm d}k \wedge
{\mathrm d}Q$ of the symplectic form to the coordinates $(k,Q)$, is determined. From
(\ref{eq:taukQ}), using a similar technique as for the computation of $a$, one finds at
lowest order
\begin{equation}
\label{eq:taukQ1}
\tau_{kQ} = \frac{ \kappa }{ \gamma } + \mathcal{O} \left(\kappa ^2 \right).
\end{equation}
Then, the effective Hamiltonian
\begin{eqnarray}
\Heff & = & \frac{ 1 }{ T } \int _0 ^T \sum _n \left[ J _n ^2 - J _n J _{n+1} + W( \rho _n )
\right] {\mathrm d}t \nonumber \\
 & = & \frac{ \kappa }{ 2\gamma } (3 - \cos k) + \mathcal{O} \left(\kappa ^2 \right) + \Veff
(Q)
\label{eq:HeffTDB}
\end{eqnarray}
is computed. The effective potential $\Veff$ defines a Peierls-Nabarro potential for
traveling DB~\cite{AMS,MS}. It is a periodic function of period $1$ whose first Fourier term
reads $V_0 \cos (2 \pi Q) $, with
\begin{equation}
\label{eq:Veff}
V_0 = \frac{ \pi ^2 }{ \gamma \sinh \left( \frac{ \pi ^2 }{ \kappa }\right)},
\end{equation}
and the so-called Peierls-Nabarro barrier is defined by $2 V_0 $. As discussed above for
Eq.~(\ref{eq:Pois1}), for values of $ \kappa $ consistent with our hypothesis $ \kappa^2 \ll
1$, such a term is exponentially small and therefore negligible.

\begin{figure}[tb]
\vspace{0cm}
\pdfBoxd{6cm}{5.2cm}{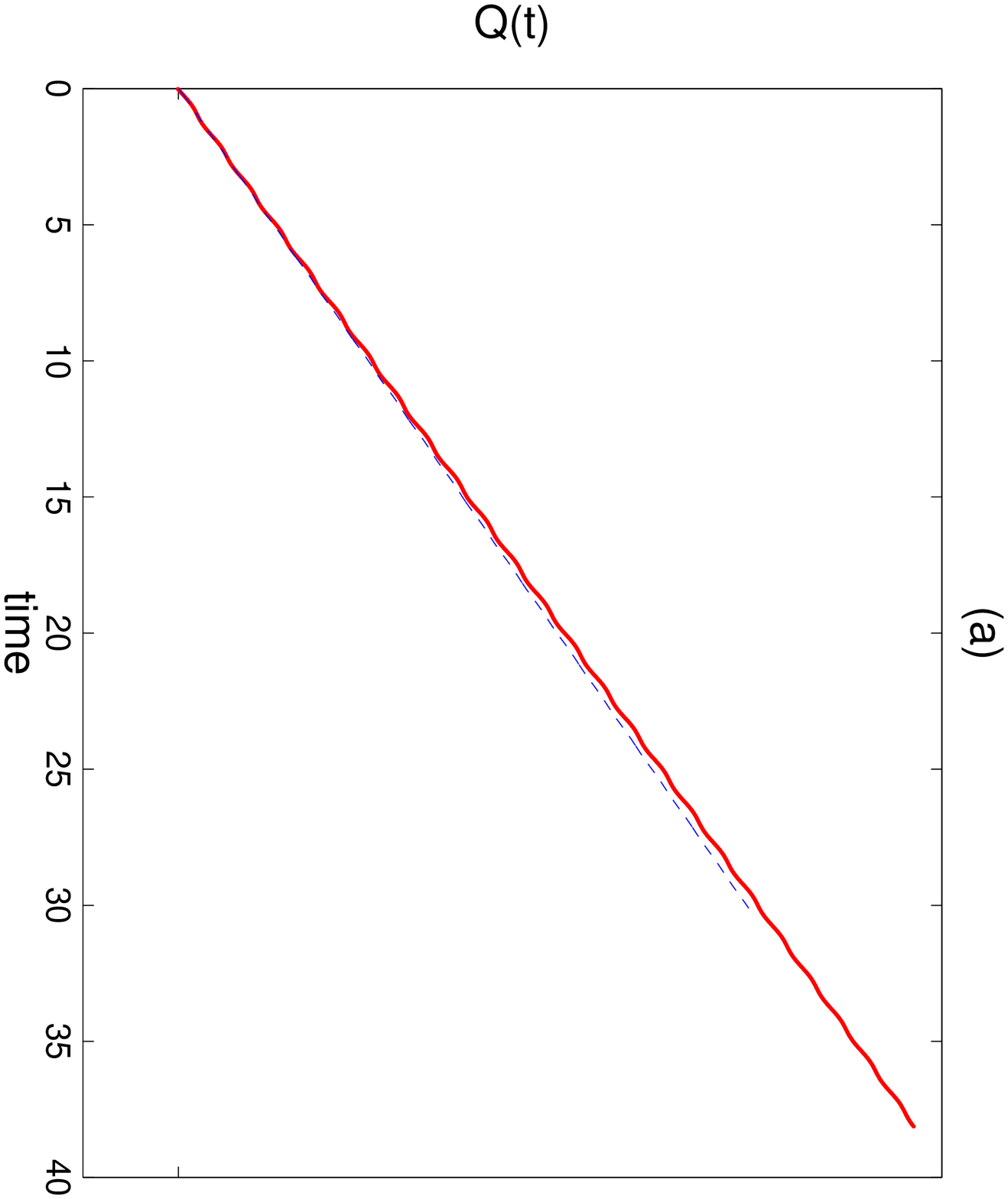}{6cm}{5.2cm}{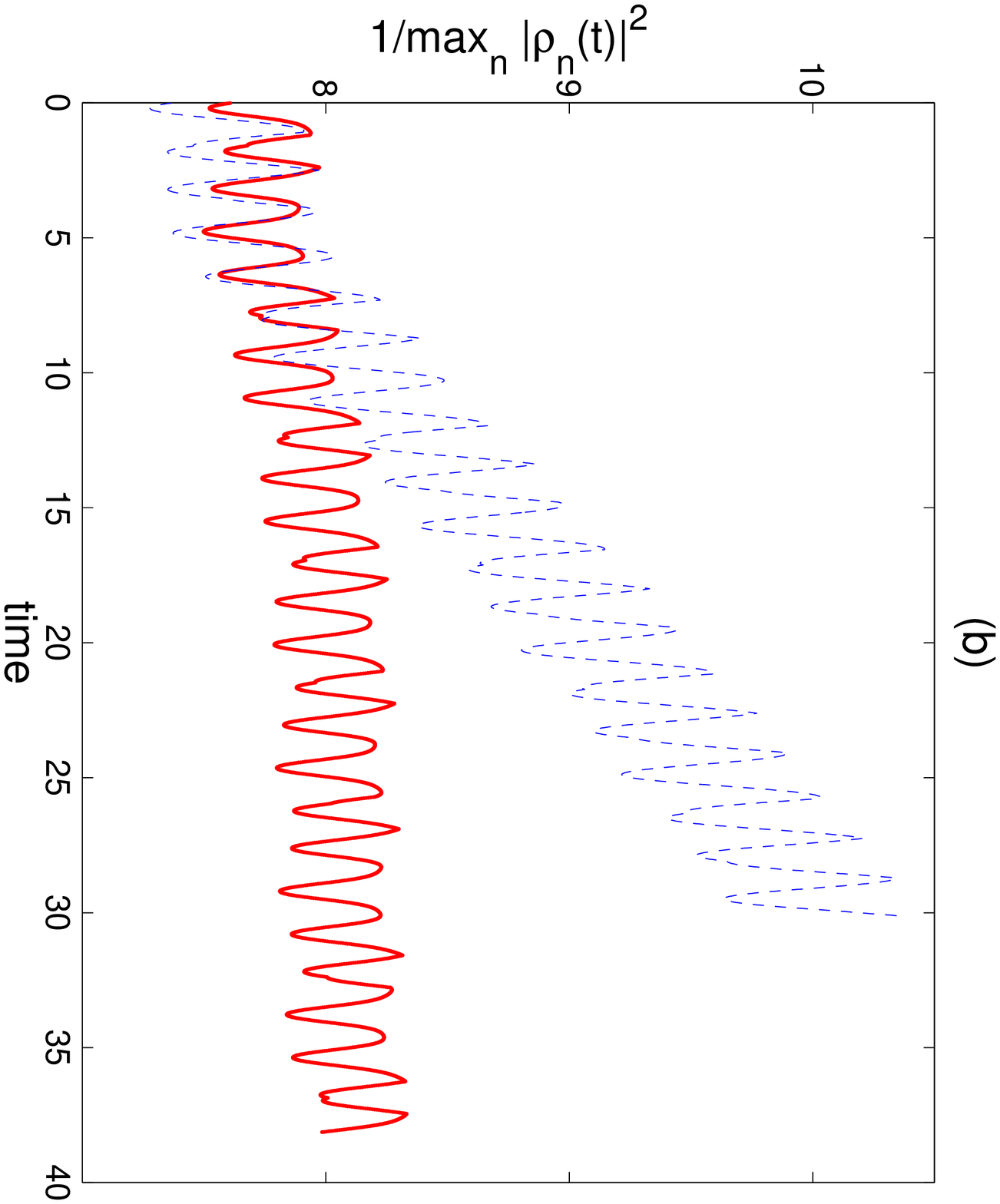}{0ex}{-0cm}{90}

\caption{\label{Fig:3} (a) Numerical integration of the equations of motion of the FPU chain
with initial condition (\ref{eq:analytic_yn}) and $\kappa=0.26, \beta =0.5$ and $k=\pi-0.4$.
A DB (solid) travels with a speed comparable to, but slightly larger than that of the
corresponding linear wave packet (dashed). (b) The (inverse of the) amplitude of a traveling
DB (solid) is oscillating around an approximately constant value, whereas the inverse of the
amplitude of a linear wave packet (dashed) clearly increases linearly in time. }
\end{figure}

The velocity $\dot{Q}$ of the DB can be estimated from the effective equations of
motion~(\ref{eq:effdyn}), finding
\begin{equation}
\label{eq:Qdot}
\dot{Q} = \frac{1 }{ \tau _{kQ}} \frac{\partial \Heff}{\partial k} \simeq \frac{ \sin k }{ 2
} + \mathcal{O} \left(\kappa  \right).
\end{equation}
Let us compare this result with the group velocity of a wave packet with wavenumber $k$,
\begin{equation}
\label{eq:domega}
\frac{{\mathrm d} \omega }{{\mathrm d}k} = \cos \left(\frac{ k }{ 2} \right),
\end{equation}
following from the dispersion relation of the linear chain $ \omega (k) = 2 \sin \left(\frac{
k }{ 2} \right)$. Therefore, assuming $|\pi - k | \ll 1 $, (\ref{eq:Qdot}) and
(\ref{eq:domega}) are equal to first order in $|\pi - k |$, and one can write approximately
\begin{equation}
\label{eq:Qdot_approx}
 \dot{Q} \simeq \frac{{\mathrm d} \omega }{{\mathrm d}k}.
\end{equation}
In conclusion, the velocity of the approximate traveling DB given by~(\ref{eq:analytic_yn})
is very close to the group velocity of the corresponding linear wave packet. Let us note that
this observation was already reported in~\cite{Ts}, although in this reference
Eq.~(\ref{eq:Qdot_approx}) was not deduced, but rather taken as an assumption. The result is
also confirmed by numerical integration of the equations of motion, illustrated in
Figure~\ref{Fig:3}(a).

Let us remark that Ansatz~(\ref{eq:analytic_yn}) does not behave like a linear wave packet.
Firstly, it has already been mentioned that this Ansatz has a singular behavior in the linear
limit $\gamma \rightarrow 0$.  Secondly, the maximum amplitude of a true linear wave packet
would evolve in time like $|\rho _n | \propto t^{-1/2} $. For the weakly localized DB
(\ref{eq:analytic_yn}), however, the amplitude is approximately proportional to the area $a$,
and is therefore predicted to be approximately constant in time. This difference in the time
evolution is confirmed by numerical simulation. In Figure~\ref{Fig:3}(b), the time evolution
of the (inverse of the) DB amplitude $ 1/ \max _n |\rho _n |^2 $ is plotted for two cases,
namely with and without the presence of a nonlinearity, showing, respectively, a decreasing
and a constant (in time) amplitude.

\section{Strongly localized discrete breathers (type III)}
\label{sec:strong}

In this section, we will not explicitly compute an effective Hamiltonian, although this
should be possible. Instead, we will restrict ourselves to estimating the Peierls-Nabarro
barrier directly from the properties of static (=non-traveling) DB.

\subsection{Analytic estimate}

Two kinds of strongly localized DB (called ``type III" in Figure~\ref{f1}) were found in the
early numerical simulations of intrinsic localized modes of anharmonic lattices. Their shapes
were approximately represented as
\begin{equation}\label{eq:PandST}
\textstyle u _n ^{\text{\tiny P}} = A \big( \cdots, 0, - \frac{ \sqrt{ 3} }{ 2},\frac{ \sqrt{ 3} }{ 2} ,
0, \cdots \big) \qquad\mbox{and}\qquad u _n ^{\text{\tiny ST}} = A \big( \cdots, 0, - \frac{1}{2} , 1,
-\frac{1}{2} , 0, \cdots \big),
\end{equation}
known respectively as the Page mode (P), which is stable, and the Sievers-Takeno mode (ST), which is unstable. The
approximations (\ref{eq:PandST}) are used for large $A^2 \omega_0 / \beta $ \cite{Ki}, and
thus are valid in a regime different from the one for which Ansatz~(\ref{eq:analytic_yn})
holds. As an family of DB, interpolating between the P and the ST mode, we use a compact
approximation (proposed by Kivshar~\cite{Ki} in a different context), namely
\begin{equation}
\label{eq:compDB}
x _n (t) = (-1) ^n A\, u(n-Q) f(\omega t),
\end{equation}
where $u$ is a truncated cosine defined by
\begin{equation}
\label{eq:truncos}
u (n) = \left\{\begin{array}{c@{\quad}l} \cos \left( \frac{ \pi }{ 3 } n\right) & \mbox{if\ }
| n | < \frac{ 3 }{ 2 }, \\ 0 & \mbox{otherwise,}\end{array}\right.
\end{equation}
and $f$ is a Jacobi elliptic function, e.g., $f( \omega t) = \mbox{\rm cn} ( \omega t ;
\kappa ) $ (modulus $\kappa = \sqrt{ 2 }$ in~\cite{Ki}). Although it is known that the FPU
system does not support compact solutions \cite{Flach}, a convergence towards a compact
function $u$ [albeit not exactly towards the truncated cosine (\ref{eq:truncos})] is observed
in the large amplitude limit (= the limit of strong localization). Therefore,
(\ref{eq:compDB}) is used here as a sketchy approximation from which the, likewise sketchy,
ST and the P modes (\ref{eq:PandST}) are recovered for integer and half-integer values of
$Q$, respectively. Equation~(\ref{eq:compDB}) defines a family of approximate DB, complying
with the requirements of (\ref{eq:famDB}), and, following the procedure of
Section~\ref{sec:proc}, it could be used to construct an effective Hamiltonian. Instead of
doing so, we will proceed straight away to the estimation of the Peierls-Nabarro barrier from
properties of the static DB, extending the results presented in \cite{S}.

Let us first estimate the area of each member of the family~(\ref{eq:compDB}). A simple
substitution of this Ansatz in~(\ref{eq:areaDB}) yields
\begin{equation}
\label{eq:areaMode}
a_Q = 2 \pi \omega_Q A_Q ^2 \langle (f ^{\prime}) ^2 \rangle \sum_{n} u ^2 (n-Q),
\end{equation}
where $Q = 0$ or $Q=\frac{1}{2} $, and $\langle \cdot \rangle$ denotes the mean value over
one period. Inserting (\ref{eq:truncos}), $ \sum_{n} u ^2 (n-Q) = \frac{ 3 }{ 2 } $ for both
the P and the ST mode. Therefore, the two modes have same area if
\begin{equation}
\label{eq:samearea}
\omega _{\text{\tiny P}} (A _{\text{\tiny P}} ) A _{\text{\tiny P}} ^2 = \omega _{\text{\tiny ST}} (A _{\text{\tiny ST}} ) A _{\text{\tiny ST}} ^2
\end{equation}
(where the symbols $ST$ and $P$ represent, respectively, any integer or half-integer value
$Q$). This equation is equivalent to saying that $A_{\text{\tiny P}}$ is a function of $A_{\text{\tiny ST}}$, e.g.\ $A_{\text{\tiny P}}
= {\mathcal A} (A_{\text{\tiny ST}})$, and the function ${\mathcal A}$ can be evaluated at least
numerically. Let us denote the corresponding periods $T_{\text{\tiny ST}} = 2\pi / \omega _{\text{\tiny ST}} (A _{\text{\tiny ST}}
)$ and $T_{\text{\tiny P}} = 2\pi /\omega _{\text{\tiny P}} (A _{\text{\tiny P}} )$. Then the Peierls-Nabarro barrier can be computed as
the difference between the energies
\begin{equation}
\label{eq:Hst}
E_{\text{\tiny ST}} = \frac{ 1 }{ T_{\text{\tiny ST}} } \int _0 ^{T_{\text{\tiny ST}}} H \big(x ^{(\text{\tiny ST})} \big) = \frac{ a }{ 2T_{\text{\tiny ST}}}
+ A _{\text{\tiny ST}} ^2 \bigg( \frac{ 5 }{ 2}\omega_0^2 \langle f^2 \rangle + A _{\text{\tiny ST}} ^2 \frac{ 82 }{ 32
} \beta \langle f^4 \rangle\bigg)
\end{equation}
and
\begin{equation}
\label{eq:Hp}
E_{\text{\tiny P}} = \frac{ 1 }{ T_{\text{\tiny P}} } \int _0 ^{T_{\text{\tiny P}}} H \big(x ^{(\text{\tiny P})} \big) = \frac{ a }{ 2T_{\text{\tiny P}}} + A _{\text{\tiny P}}^2
\bigg( \frac{ 9 }{ 4}\omega_0^2 \langle f^2 \rangle + A_{\text{\tiny P}} \frac{\sqrt{27}}{4} \alpha \langle
f^3 \rangle+ A _{\text{\tiny P}} ^2 \frac{ 81 }{ 32 } \beta \langle f^4 \rangle\bigg).
\end{equation}

\subsection{Numerical results}

\begin{figure}[htb]
\centerline {
\includegraphics[width=8.5cm,angle=270]{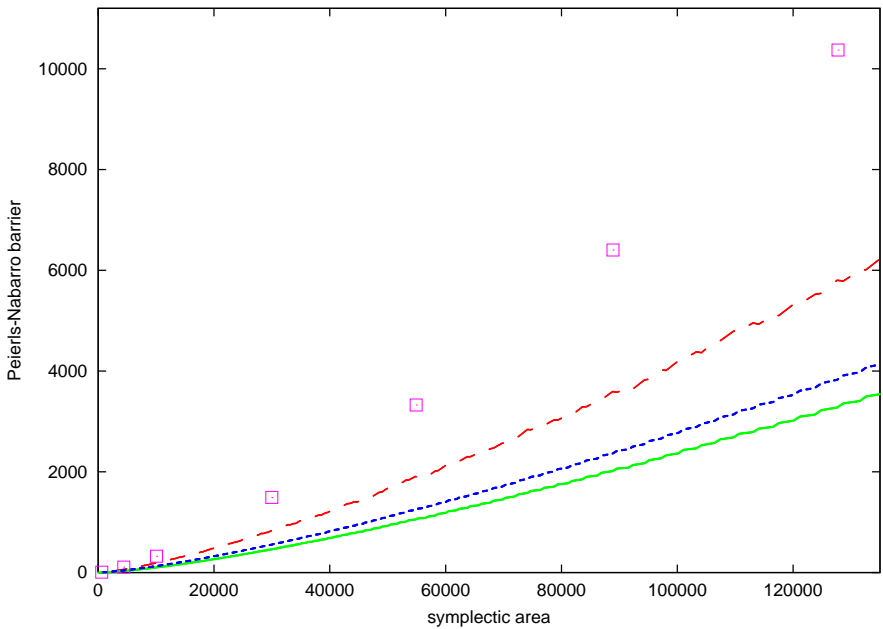}
}
\caption{ \label{PNbarrier} {\protect\small
Different approximations for the Peierls-Nabarro barrier $\Delta E_{\text{\tiny PN}}$ versus the
symplectic area $a$. Dashed: numerically exact Peierls-Nabarro barrier, solid: approximation
(2.), dotted: ``kinetic'' approximation (3.), points: upper bound on the depinning energy
(see text for details).
}}
\end{figure}

Numerical simulations were performed in order to check the validity of the
approximations~(\ref{eq:Hst}) and (\ref{eq:Hp}), and hence of the underlying
assumptions~(\ref{eq:compDB}) and (\ref{eq:truncos}). As parameter values, we chose
$\omega_0=1$, $\alpha=0$, and $\beta=1$ in the interaction potential (\ref{eq:W}). Discrete
breathers can be calculated numerically up to machine precision, for example by means of a
continuation of periodic orbits from the anticontinuous limit~\cite{MA}. We have computed
four different quantities related to the mobility of a discrete breather, which are compared
in Figure~\ref{PNbarrier}.
\begin{enumerate}
\item Numerically exact Peierls-Nabarro barrier: Computing ST and P modes with the same value
of the symplectic area $a$, the numerically exact Peierls-Nabarro barrier $\Delta
E_{\text{\tiny PN}}=|E_{\text{\tiny P}}-E_{\text{\tiny ST}}|$ can be obtained as the difference in energy between the two modes
(dashed in Figure~\ref{PNbarrier}).
\item An approximation to the Peierls-Nabarro barrier: Equations~(\ref{eq:Hst}) and
(\ref{eq:Hp}) allow to approximate the Peierls-Nabarro barrier by computing the difference
between their right hand sides (solid in Figure~\ref{PNbarrier}). The values $\langle f^2
\rangle =\frac{1}{2}$, $\langle f^3 \rangle = 0$, and $\langle f^4 \rangle = \frac{3}{8}$
have been used, which are found when $f$ is assumed to be a cosine function. The values of
the oscillation periods $T_{\text{\tiny ST}}$ and $T_{\text{\tiny P}}$, as well as the (maximum) DB amplitudes $A_{\text{\tiny ST}}$
and $A_{\text{\tiny P}}$, have been extracted from the numerical data.
\item A second approximation to the Peierls-Nabarro barrier: Computing discrete breathers
numerically, one notices that the oscillation periods $T_{\text{\tiny ST}}$ and $T_{\text{\tiny P}}$ corresponding to the
same value of the symplectic area $a$ are clearly different, and $T_{\text{\tiny ST}} > T_{\text{\tiny P}}$. In fact, it
is observed that the main contribution to the above approximation (2.) of the Peierls-Nabarro
barrier $\Delta E_{\text{\tiny PN}}$ stems from the kinetic part,
\begin{equation}
\Delta E_{\text{\tiny PN}} \approx \frac{a}{2}\left(\frac{1}{T_{\text{\tiny P}}}-\frac{1}{T_{\text{\tiny ST}}}\right).
\end{equation}
This quantity is used as a second approximation of the Peierls-Nabarro barrier (dotted in
Figure~\ref{PNbarrier}).
\item An upper bound on the depinning energy: The aim of the computation of the
Peierls-Nabarro barrier is to obtain an estimate of the energy necessary to ``depin'' a
discrete breather (i.e., to kick a static DB such that it starts traveling along the
lattice). An upper bound on the depinning energy (points in Figure~\ref{PNbarrier}) can be
obtained from a simple numerical experiment which consists in kicking a static DB with a
certain momentum distribution, make it evolve in time by means of numerical integration of
the equations of motion, and see if it starts traveling along the lattice. This procedure, of
course, depends strongly on the momentum distribution applied, and a rough numerical
optimization scheme was employed in order to approximate the minimum energy necessary for
depinning.
\end{enumerate}
%
Since Ansatz (\ref{eq:compDB}) is supposed to be valid in the limit of large
$A^2\omega_0/\beta$ ($=A^2$ with our choice of parameters), large values of the DB amplitude
(and hence the symplectic area $a$) have been considered. However, the quality of the
approximations does not change significantly with increasing $a$, and the relative error
between exact result and approximation seems to be more or less constant. The quality of the
approximations is reasonable, albeit not excellent, differing from the exact Peierls-Nabarro
barrier roughly by a factor of two.

When computing numerically the Peierls-Nabarro barrier $\Delta E_{\text{\tiny PN}}(a)=|E_{\text{\tiny P}}(a)-E_{\text{\tiny ST}}(a)|$
for some value of the symplectic area $a$, we were surprised to find $E_{\text{\tiny P}}(a)>E_{\text{\tiny ST}}(a)$.
Typically, a stable mode is expected to correspond to the minima of the Peierls-Nabarro
potential, and an unstable mode to the maxima. Finding the reverse situation, we conclude
that the motion of a traveling DB in the FPU system is effectively described by the dynamics
of a particle with negative effective mass in the Peierls-Nabarro potential.

The upper bound on the depinning energy we found in the numerical experiment differs from the
Peierls-Nabarro barrier roughly by a factor of two. A more refined optimization of the
momentum distribution employed in the depinning experiment might further improve the match.

\section{Conclusions}

Effective Hamiltonians are constructed for various types of DB which can be found in the FPU
model. These types, labelled type I, II, and III in Figure~\ref{f1}, are characterized by
their degree of spatial localization. Type I is not a true discrete breather, as its
localization depends on the system size, but it represents a localized wave which can be set
into motion. The Peierls-Nabarro barrier is found to be zero in first approximation, and the
localized wave travels at constant speed. An explicit estimate shows that this speed is not a
group velocity and depends on the system size. DB of type II are called weakly localized.
They are small in amplitude and can be analytically approximated by employing a recent result
by James \cite{J,J03}, proving existence of DB in the FPU system. In this case, the
Peierls-Nabarro barrier is found to exist, but it is negligible, being exponentially small
with respect to the DB amplitude. DB of type II are found to travel at a speed approximately
equal to the group velocity of the corresponding wave packet, but, in contrast to the wave
packet, without dispersion. This group velocity is much smaller than the sound velocity.
Finally, DB of type III are considered, being strongly localized and having an amplitude
which is not small. The well-known Page and Sievers-Takeno modes are of this type, and
numerically they have frequently been observed in a traveling state. It should be possible to
provide an analytic estimate of their traveling speed by constructing an effective
Hamiltonian, following the procedure of Section~\ref{sec:proc} and making use of
Ansatz~(\ref{eq:compDB}), but this is yet to be done. The important feature which
distinguishes DB of type III from those of type II is that their Peierls-Nabarro barrier is
non-negligible. This barrier is estimated without explicitly computing the effective
Hamiltonian, and the estimate is in reasonable agreement with the energy necessary to render
a DB mobile in a numerical experiment.

\section*{Acknowledgments}

J.-A.\ S.\ acknowledges R.\ S.\ MacKay for his suggestion to continue the work initiated
in~\cite{MS} in the direction opened by G.\ James. The latter is acknowledged for interesting
discussions. Part of this work was done during M.\ K.'s stay at the Universit\`a di Firenze, Italy, in the group of
Roberto Livi. Work supported in part by the European Community's Human Potential Program under
contract HPRN-CT-1999-00163, LOCNET.

\medskip



\end{document}